\documentclass[conference]{IEEEtran}

\usepackage{cite}
\usepackage{amsmath,amssymb,amsfonts}
\usepackage{algorithmic}
\usepackage{graphicx}
\usepackage{textcomp}
\usepackage{xcolor}
\usepackage{caption}

\usepackage{listings} 

\begin{document}

\title{Helm --- What It Can Do and Where Is It Going?\\
}

\author{\IEEEauthorblockN{Michael Howard}
\IEEEauthorblockA{\textit{Computer Science Department} \\
\textit{Portland State University}\\
Portland, USA \\
mihoward@pdx.edu}
}

\maketitle

\begin{abstract} \label{sec:abs}
Deploying an application into a Kubernetes cluster requires sending a manifest file to the cluster's control plane interface.  This action is typically performed through a \emph{kubectl} client which is configured and authorized to communicate with the control plane's Uniform Resource Locator (URL). An application typically requires many Kubernetes resources such as pods, deployments, secrets, service and volumes.  Configuring each of these through manifest files requires complex scripting, especially when there are numerous resources needed.

A solution to the complex management tasks is Helm.  Helm provides both a tool and underlying framework that packages the necessary manifest files.  These packages are deployed through a single step install command which abstracts all the underlying control plane interaction from the user.  Similar to application installs through Debian's package manager \emph{dpkg}, packages are shared through local and remote repositories and allow the user to easily install, update, delete or handle concurrent versions. 
\end{abstract}


\section{Introduction} \label{sec:intro}
Helm is a package manager for Kubernetes applications that originally began as a Deis project in 2015 and as introduced at the inaugural KubeCon.  It was accepted as an open-source project to Cloud Native Computing Foundation (CNCF) in 2016 and achieved ``graduated'' status in 2020.  The packages are called \emph{charts}.  A chart is a bundle of information necessary to create a Kubernetes application.  \cite{helmdocs}

This paper discusses the benefits and high-level architecture of Helm.  Section \ref{sec:what} addresses what a user can do with Helm while Section \ref{sec:components} digs into the executable components and the duties for each.  Section \ref{sec:charts} expands the composition of a chart as well as the convenient template support allowing the dynamic generation of required manifest files.  Next, Section \ref{sec:repsitories} details how the charts are stored in repositories and synchronized with remote instances.  Upon install, each chart creates the necessary Kubernetes resources in a preset order.  This order is listed in Section \ref{sec:installed}.  Finally, Section \ref{sec:workflows} digs further into workflows a user may follow, the security features and a convenient post-render script option to avoid the need to fork and modify an existing chart. 

\section{What Does Helm Do?} \label{sec:what}
Helm operates on three object types: charts, repositories and releases.  A chart is a Helm package containing all resource definitions necessary to run an application inside a Kubernetes cluster.  Charts are collected and shared within a \emph{repository}.  A \emph{release} is an instance of a chart running in a Kubernetes cluster.  Since a chart can be installed multiple times within a cluster, each will have its own release with a unique release name.

The Helm tool allows the user to do the following:
\begin{itemize}
    \item Create a new chart.
    \item Archive a chart into a .tgz package.
    \item Interact with a chart repository.
    \item Install and uninstall a chart into a Kubernetes cluster.
    \item Manage the release cycle of a chart.
\end{itemize}

Helm is a pluggable architecture and a current list of plugin projects is hosted on the official website\footnote{https://helm.sh/docs/community/related/\#helm-plugins}.

\section{Executable Components} \label{sec:components}
Helm is an executable that is implemented in two parts.  The first is the command-line client called the \emph{Helm Client} and provides an interface for the end users.  The Helm Client enables the following functionality: \cite{helmdocs}
\begin{itemize}
    \item Chart development.
    \item Managing the chart repositories.
    \item Managing releases.
    \item Interfacing with the Helm Library.
\end{itemize}

The second part is the \emph{Helm Library} which contains the core logic for all Helm operations.  The Library interfaces with the Kubernetes Application Programming Interface (API) server and provides the following capabilities:
\begin{itemize}
    \item Combining a chart with configuration to build a release.
    \item Installing a chart into Kubernetes cluster.
    \item Upgrading and uninstalling charts in the Kubernetes cluster.
\end{itemize}

The Helm Library is a standalone component and can be directly accessed by different clients.  Both the client and library are implemented with the Golang programming language.  The library communicates with the Kubernetes cluster using JavaScript Object Notation (JSON) messages via a Representational State Transfer (REST) API.

\section{Charts} \label{sec:charts}
\begin{figure}[htbp]
    \centerline{\includegraphics[width=.9\linewidth, keepaspectratio]{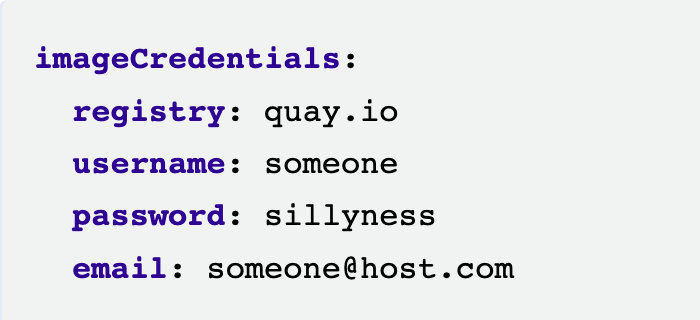}}
    \captionsetup{width=.8\linewidth}
    \caption{Source YAML to feed the template.}
    \label{fig:source_yaml}
\end{figure}

\begin{figure}[htbp]
    \centerline{\includegraphics[width=.9\linewidth, keepaspectratio]{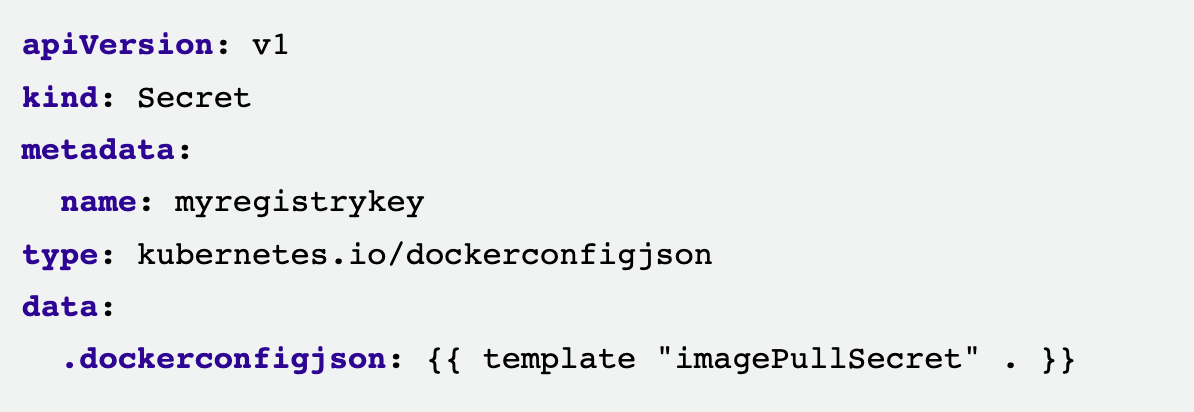}}
    \captionsetup{width=.8\linewidth}
    \caption{Resulting YAML.}
    \label{fig:dest_yaml}
\end{figure}

Helm uses a chart object to capture all resource files needed to deploy an application into a Kubernetes cluster.  The chart is built using Go templates\footnote{https://pkg.go.dev/text/template?utm\_source=godoc} which are further enhanced through Sprig library\footnote{https://masterminds.github.io/sprig/} functions.  For example, to generate a Kubernetes manifest files as seen in Figure \ref{fig:dest_yaml}, a source Yet Another Markup Language (YAML) file, as in Figure \ref{fig:source_yaml}, is used as an input to the Go template in Listing \ref{lst:template}.  The template is called directly within the destination manifest file.

\lstset{captionpos=b, basewidth=0.5em, flexiblecolumns=true, basicstyle=\scriptsize,
 showspaces=false, xleftmargin=.5in, xrightmargin=.5in, backgroundcolor=\color{lightgray!20}}
\begin{lstlisting}[language={Go}, caption={Go template.}, label=lst:template]
{{- define "imagePullSecret" }}
{{- with .Values.imageCredentials }}
{{
- printf "{\"auths\":{\"%s\":
{\"username\":\"%s\",\"password\":\"%s\",
\"email\":\"%s\",\"auth\":\"%s\"}}}"
.registry .username .password 
.email (printf "%s:%s" .username
.password | b64enc) | b64enc 
}}
{{- end }}
{{- end }}
\end{lstlisting}

\begin{figure}[htbp]
    \centerline{\includegraphics[width=.9\linewidth, keepaspectratio]{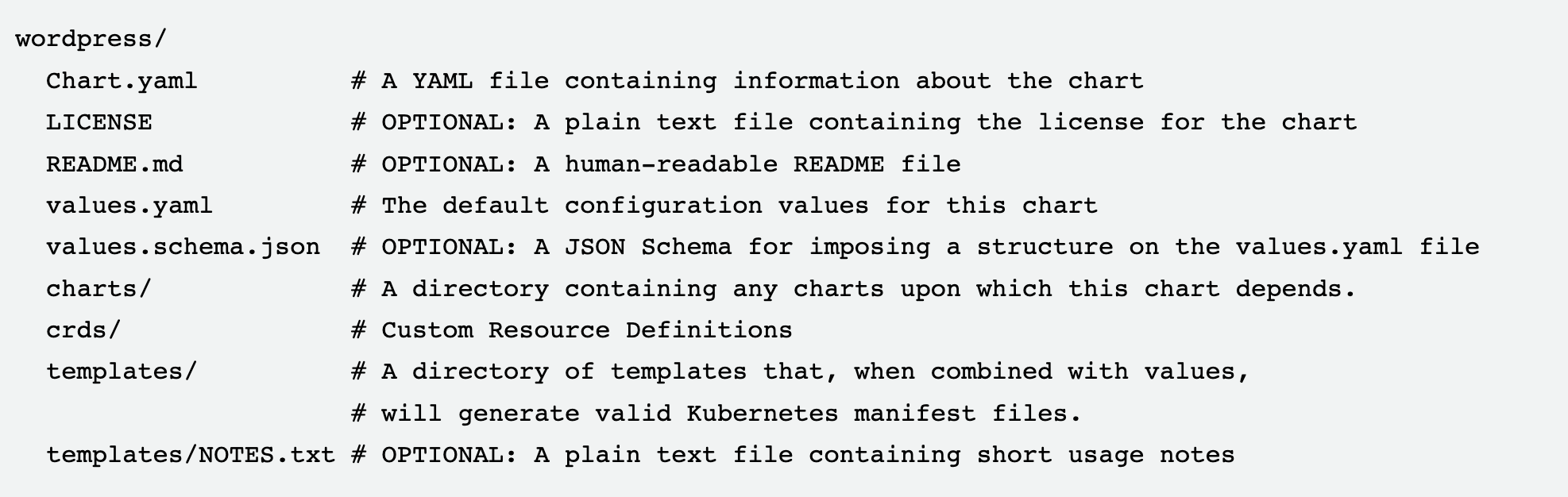}}
    \captionsetup{width=.8\linewidth}
    \caption{Files in a chart bundle.}
    \label{fig:chart_structure}
\end{figure}

The initial directory structure for the chart is created through the \emph{helm create} command.  YAML and Go template files are then modified within that directory structure and packaged using \emph{helm package}.  The resulting .tgz bundle is now ready to be added to a local Helm repository.  Within the .tgz bundle, chart files are organized as seen in Figure \ref{fig:chart_structure}.

In chart.yaml, dependencies are optionally specified.  This allows the installation of the chart to be restricted unless the other dependent charts are also installed.  A linter is also provided via \emph{helm lint} allowing a user to validate the existing chart file structure and its declarations.  A chart also has the capability to act as a library or helper.  This means that snippets from a chart are embedded in templates of other charts avoiding the need for the user to recreate that particular functionality.

\section{Repositories} \label{sec:repsitories}
\begin{figure}[htbp]
    \centerline{\includegraphics[width=.9\linewidth, keepaspectratio]{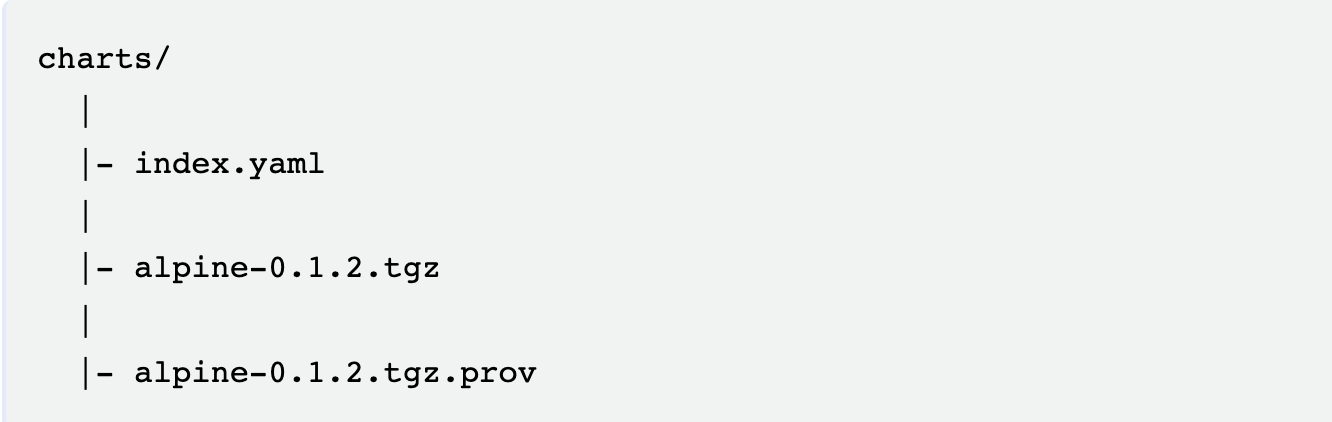}}
    \captionsetup{width=.8\linewidth}
    \caption{Directory structure of a repository.}
    \label{fig:dir_structure}
\end{figure}

The Helm repository is simply a collection of chart packages organized within a file system directory.  Figure \ref{fig:dir_structure} shows the directory structure within the repository.  

The charts themselves are stored .tgz bundles.  The index.yaml file is necessary in every repository to manage and identify the charts within.  This index is initially generated and refreshed after adding a new chart via the \emph{helm repo index} command.  Local repositories are optionally synced with remote counterparts.  Helm primarily performs all chart actions (i.e. install, list, search) via a repository.  A distributed community Helm repository is located at Artifact Hub \footnote{https://artifacthub.io/packages/search?kind=0}.  The client communicates with a remote repository via Hypertext Transport Protocol (HTTP).  Thus, any HTTP-supported cloud storage system may be used to host the repository.  Examples are Google Cloud Storage buckets, Amazon S3 buckets, GitHub Pages or a custom web server.  Secure Socket Layer (SSL) is also supported.

\section{Installed Resources} \label{sec:installed}
During a Helm install operation, the following Kubernetes resources are installed in order:
\begin{enumerate}
	\item Namespace
	\item NetworkPolicy
	\item ResourceQuota
	\item LimitRange
	\item PodSecurityPolicy
	\item PodDisruptionBudget
	\item ServiceAccount
	\item Secret
	\item SecretList
	\item ConfigMap
	\item StorageClass
	\item PersistentVolume
	\item PersistentVolumeClaim
	\item CustomResourceDefinition
	\item ClusterRole
	\item ClusterRoleList
	\item ClusterRoleBinding
	\item ClusterRoleBindingList
	\item Role
	\item RoleList
	\item RoleBinding
	\item RoleBindingList
	\item Service
	\item DaemonSet
	\item Pod
	\item ReplicationController
	\item ReplicaSet
	\item Deployment
	\item HorizontalPodAutoscaler
	\item StatefulSet
	\item Job
	\item CronJob
	\item Ingress
	\item APIService
\end{enumerate}

\section{Workflows} \label{sec:workflows}
Helm can support installing a new chart from either a local or remote repository.  A local file in .tgz format is also supported.  Configuration key-value pairs are optionally passed to the install action via a YAML-format file or directly set with the --set argument.  Once deployed as a release, it may be upgraded to enforce a change or rolled back to recover from a failure.  The helm tool can query the history of every incremental release.  An uninstall action will fully destroy the Kubernetes deployment and all associated resources.

Integrity and provenance support is provided with Helm.  Upon creating a package, a command line option provides for signing using GNU Privacy Guard (GPG) and storing the resulting string in a sidecar .prov file.  A verify option passed to the install command checks the generated signature with the .prov.  The install will fail if they do not match guaranteeing the integrity of the package.

Another flexible feature allowing for customization is the post-render option on install.  This option sends all manifests to the binary specified by the command line option.  It is the responsibility of that binary to then manipulate the manifest and return via its standard output.  Forking and modifying charts are avoided through using this feature. 

\section{Conclusion and Discussion} \label{sec:conclusion}
This paper begins with the discussion of the problem of managing application deployments into a Kubernetes cluster.  The complex interactions with the Kubernetes control plane involves multiple commands each requiring extensive arguments.  In addition, the Kubernetes resources themselves must be specified in manifest files.  Thus, for any reasonably complex deployment, a user must maintain the manifests as well as script files that can handle the order of the resources and different actions such as install, delete, upgrade and multi-version install.

A solution to this problem is a package manager such as Helm.  Helm allows the manifest files for a given application to be bundled into a chart package and stored in a repository.  Repositories may easily sync between multiple instances and therefore allow public and private replicas.  A Helm Command Line Interface (CLI) tool provides a single command to easily install a chart from a specified repository to a destination Kubernetes cluster.  This command abstracts all the complexity, manifests and multiple steps involved with control plane interaction.  Once a chart is installed, the application is deployed in the cluster as a release.  Multiple releases may exist simultaneously and are easily updated or deleted through the Helm command set.

The Helm charts also support templates.  The template directives allow for a dynamic generation of the manifest files which are necessary since each chart install will have unique properties that must be integrated into the corresponding manifests sent to the cluster.  When generating a new chart package, there is a signature option based on GNU Privacy Guard that guarantees the source and integrity of the package.  If modifications are needed when installing an existing chart, it is not always necessary to fork and modify.  Helm provides a post-render script option allowing the user to modify the manifest files before they are sent to the control plane.  Even the Helm tool itself is not static, rather it supports a plugin architecture allowing a user to create Helm extensions in any chosen programming language.

Helm is a Cloud Native Computing Foundation (CNCF) project and achieved graduated status as of 2020.  At the time of this writing, the project has released version 3.8.2.  Version 3.9.0 is scheduled through GitHub \footnote{https://github.com/helm/helm/milestones} and uses milestones (containing a set of issues) to define the scope.  Only the next version is present.  The project maintainers do not appear to have an extended roadmap available to see where this project is going in the future.  Such a roadmap would be helpful, but it is acknowledged that Helm is an interface into Kubernetes and thus must be more responsive in nature.  Helm offers great simplification for any complex Kubernetes deployment and is an excellent tool to use in any Continuous Integration Continuous Deployment (CI/CD) pipeline.

\section{Acknowledgements} \label{sec:ack}
The author would like to acknowledge the contributions of Dr. Wu-Chang Feng at Portland State University.  His advisorship, funding and support made this research possible.

\vspace{12pt}

\end{document}